\documentclass[noshowpacs,amsmath,twocolumn,superscriptaddress,10pt,aps,prb]{revtex4-1} % this article uses the Revtex 4-1 format
\usepackage{setspace}
\usepackage{amsmath}
\usepackage{breqn}
\usepackage{graphicx}
\usepackage[nearskip,margin = 0pt,caption=false]{subfig}

\usepackage{verbatim}
\usepackage{amsfonts}
\usepackage{amssymb}
\usepackage{epstopdf}
\usepackage{lipsum}
\usepackage{siunitx}
\usepackage{xcolor}
\usepackage{array}
\usepackage[normalem]{ulem}
\usepackage[resetlabels,labeled]{multibib}
\DeclareGraphicsExtensions{.pdf,.eps,.png,.jpg,.mps}
\usepackage{etoolbox}

\begin{document}
\title{Augmenting On-Chip Microresonator through Photonic Inverse Design}
% \title{Photonic inverse design of integrated silicon resonators}
\author{Geun Ho Ahn$^{1}$, Ki Youl Yang$^{1}$, Rahul Trivedi$^{1,2}$, Alexander D. White$^{1}$, Logan Su$^{1}$, Jinhie Skarda$^{1}$, Jelena Vu\v{c}kovi\'{c}$^{1,\dagger}$\\
\vspace{+0.05 in}
$^1$E. L. Ginzton Laboratory, Stanford University, Stanford, CA 94305, USA.\\
$^2$Max-Planck-Institut f\"{u}r Quantenoptik, Hans-Kopfermann-Str. 1, 85748 Garching, Germany.\\
{\small $^{\dagger}$Corresponding author: jela@stanford.edu }}

\begin{abstract}
    \noindent Recent advances in the design and fabrication of on-chip optical microresonators has greatly expanded their applications in photonics, enabling metrology, communications, and on-chip lasers. Designs for these applications require fine control of dispersion, bandwidth and high optical quality factors. Co-engineering these figures of merit remains a significant technological challenge due to design strategies being largely limited to analytical tuning of cross-sectional geometry. Here, we show that photonic inverse-design facilitates and expands the functionality of on-chip microresonators; theoretically and experimentally demonstrating flexible dispersion engineering, quality factor beyond 2 million on the silicon-on-insulator platform with single mode operation, and selective wavelength-band operation. 

\end{abstract}

\maketitle

%\begin{figure*}[t!]
%\centering
%\includegraphics[width=0.85\linewidth]{Fig1.png}
%\captionsetup{format=plain,justification=RaggedRight}
%\caption{\label{fig:Fig1}{\bf{Schematic of Proposed Techniques for Inverse Design of On-Chip Microresonator}}  Illustration showing (\textbf{a}) how on-chip microresonators - microring resonator and fabry-perot resonator- can be inverse designed. For the case of microring resonator, intracavity element can be inverse designed, and for the fabry-perot resonator, the reflectors can be inverse designed. (\textbf{b}) Illustration of intracavity partially transmitting element leading to a flexible mode splitting of the propagating modes in the resonator. Illustrations of inverse designed fabry-perot resonator for (\textbf{c}) dispersion engineering, (\textbf{d}) wavelength selective operation, and (\textbf{e}) fundamental mode selective wide width reflectors for high Q-factor operations.}
%\end{figure*}

\noindent On-chip optical microresonators are critical to a wide range of applications, including but not limited to lasers, optical communication, nonlinear optics, and quantum optics\cite{Vahala:2003:Nature,Lipson:2005:Nature,Englund:2005:PRL}. While relying on the same basic operational principles, resonator designs for different technological applications require vastly different design considerations. In particular, several important metrics, including dispersion\cite{Okawachi:2014:OL,Yang:2016:NaturePhotonics,Weiner:2017:NatureCommunications,Papp:2019:ACSPhotonics,Bowers:2020:PhotonicsResearch}, quality factor\cite{Bowers:2021:NaturePhotonics,Blumenthal:2021:NatureCommunications,Yang:2018:NaturePhotonics,Kippenberg:2018:Optica} and operation bandwidth\cite{Tang:2018:Optica,Loncar:2019:Optica,Lipson:2015:NatureCommunications}, need to be co-optimized with the specific application dictating their relative importance. Current microresonator design methodologies mostly rely on tuning a small set of design parameters, such as the resonator geometry or the material stack, guided by an analytical or numerical modelling of the device performance. Consequently, it is a challenge to hit multiple design targets. % that are often needed in a number of applications.
\\

\noindent An alternative strategy is to use \emph{inverse design} --- here, the designer first constructs a cost function, expressed in terms of the electric fields generated by the device under different excitation conditions, which captures the required design targets. A gradient based optimization algorithm can then be used to efficiently explore a large space of devices to optimize this cost function. Application of this methodology to designing traditional integrated optical devices, such as wavelength demultiplexers \cite{Piggott:2015:NaturePhotonics}, mode converters \cite{inv_mode_2016}, grating couplers \cite{Su:18} has resulted in devices that achieve high performance, high bandwidth and low cross-talk within a much smaller device footprint when compared to their traditionally designed counterparts. Here, we consider the application of inverse design for on-chip microresonator design --- we numerically and experimentally demonstrate this with three applications (i) engineering the microresonator resonances (dispersion engineering), (ii) achieving high quality (Q) factors in multi-mode waveguide-based resonators and (iii) designing wavelength-band selective resonances in microresonators.

\noindent The first design problem that we address is dispersion engineering --- several applications in nonlinear optics rely on finely tuned microresonator dispersions. For instance, the dispersion of the microresonator needs to be finely controlled for frequency comb applications, both to phase match for the initiation of optical parametric oscillation \cite{Kippenberg:2018:Science}, as well as to control the comb's spectral shape and bandwidth \cite{Kippenberg:2018:Science,Loncar:2019:Nature,Afltouni:2020:CLEO}. Traditional dispersion engineering approaches attempt to design the frequency-dependent phase accumulated by an optical field propagating in the microresonator by  either tuning the resonator cross-sectional geometry\cite{Okawachi:2014:OL,Yang:2016:NaturePhotonics}, carefully selecting a material stack\cite{Kippenberg:2012:OE} or utilizing grating structures\cite{Papp:2019:ACSPhotonics,Bowers:2020:PhotonicsResearch}. While these designs are able to initiate the nonlinear wave mixing processes, it is difficult for them to achieve finer spectral control over the nonlinear process \cite{Bell:Natpho}. Recently, more sophisticated design methods that couple multiple resonator modes have been employed to expand this design space --- these include photonic molecules \cite{Loncar:2019:NaturePhotonics, Weiner:2017:NatureCommunications, Kippenberg:2018:NatureCommunications:visiblecomb,Vahala:2017:NatureCommun:visiblecomb} and mode hybridization effects at photonic crystal edges \cite{Yu:2021:NaturePhotonics, Yu:2020:PhotonicResearch}. By posing the design of either a frequency-dependent phase or the coupling matrix between multiple resonator modes as an inverse design problem, we numerically and experimentally demonstrate a wide-range of dispersion profiles that are relevant for nonlinear optics applications.

\begin{figure*}[t!]
\centering
\includegraphics[width=\linewidth]{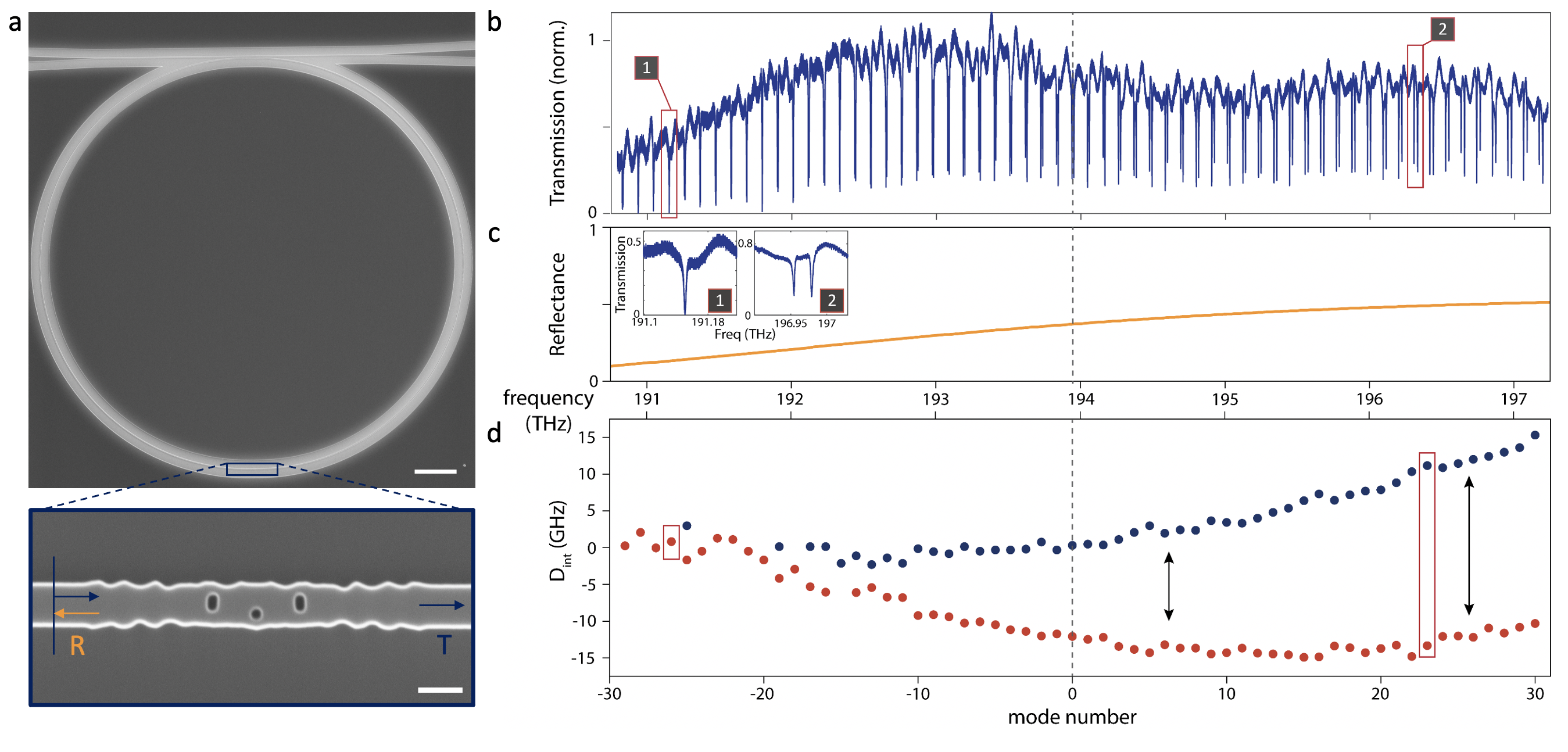}
%\captionsetup{format=plain,justification=RaggedRight}
\caption{\label{fig:Fig2}{\bf{Inverse designed intracavity element for microring resonator frequency engineering}} (\textbf{a}) SEM of fabricated Si microring resonator with the partially reflective element in the microring. The microring resonator is designed with a ring radius of 100 $\mu m$ and the width of the waveguide is 500 $nm$. The scale bar here is 20$\mu m$ and 500$nm$ respectively. (\textbf{b}) Transmission spectrum of the fabricated device with the inverse designed intracavity element. (\textbf{c}) simulated reflectance spectrum of the corresponding inverse designed intracavity element with increasing reflectance as frequency increases. (\textbf{d}) Measured integrated dispersion of the split fundamental modes, indicated by blue and red points for each side of the split modes.}
\end{figure*}

\noindent Along with achieving fine dispersion control, it is also critical in many applications for microresonators to have a high quality (Q) factor -- the next design challenge we address. A high Q factor is crucial for the resonant enhancement of lasing\cite{Bowers:2021:NaturePhotonics}, nonlinear wave mixing \cite{Srinivasan:2019:NaturePhotonics}, light-matter interaction\cite{Dirk_PhysRevLett} and narrow band filters for on-chip communication\cite{Watts:2012:OL,Watts:2014:NatureCommunications,Ram:2018:Nature}. Apart from material absorption, the Q-factor of the resonator modes are limited by the microresonator surface roughness. Consequently, one of the key strategies for improving the quality factor is to optimize the fabrication process to improve surface roughness to reduce the scattering losses\cite{Lipson:2015:NatureCommunications, Watts:2012:OL}. Alternatively, the interaction of the resonator mode with the rough resonator surface can be reduced by using wider-width waveguides to form the microresonator\cite{Watts:2012:OL}. However, wider-width waveguides in general support multiple propagating modes, which can couple to the fundamental resonator mode and adversely impact its Q-factor. An adiabatic taper can be used to suppress higher order modes \cite{Kordts:16}, but while this is an effective way of achieving high quality factor single mode operation, the adiabatic taper makes the resonator very long, limiting applications to those that can tolerate a small free-spectral range (FSR). In this paper, we demonstrate that using inverse design, the coupling of the fundamental resonator mode to higher order waveguide modes can be prohibited within a much shorter device area, allowing for flexible engineering of the microresonator FSR and Q-factor.

\noindent Finally, we consider the problem of designing microresonators that support resonant modes at two distinct and well-separated resonant bands. Such wavelength selective band operation is critical for nonlinear interaction between optical wavelengths that are far apart while preventing leakage into undesired bands \cite{mckenna2021ultralowpower} which could significantly reduce the efficiency of the nonlinear process. Traditional resonator designs find it difficult to implement such frequency-selective resonance formations while also maintaining high Q-factor and dispersion control. We resolve this problem by posing the frequency selectivity of the resonator as an inverse design problem, and numerically and experimentally demonstrate wavelength selective band operation.\\

% With all the important metrics of microresonators in consideration, there is still a lack of comprehensive design tools to explore the full gamut of microresonator designs. Most current techniques are limited to analytical theory to guide the design process. Not only do these techniques explore a small design space, this design space becomes even more limited when considering scalable fabrication (e.g.~foundry fabrication\cite{GF:2020OFC,Piggott:2020:ACSPhotonics}), wherein several parameters such as the thickness of the optical film and available optical materials are fixed. Hence, achieving the required design targets becomes an even more challenging problem. 

% In this article, we propose and experimentally validate a new strategy for designing microresonators using photonic inverse design\cite{Piggott:2015:NaturePhotonics,Rodriguez:2018:NaturePhotonics} that opens up the feasible design space allows for engineering the microresonators with practical design constraints. We design on-chip microresonators for a variety of applications relevant to linear and nonlinear optics. First, we engineer microresonator dispersion through two optimization approaches: mode splitting engineering and phase compensation, as shown schematically in FIG1 b and FIG1 c. In the second section, we present a wide waveguide resonator design optimized for a high quality factor and single mode operation (FIG1 e). Finally, we introduce resonators with inverse designed reflectors that allow for a selective wavelength band operation (FIG1 d). \\

\noindent{\bf Inverse Design of Microresonator Dispersion.}

\begin{figure*}[t!]
\centering
\includegraphics[width=\linewidth]{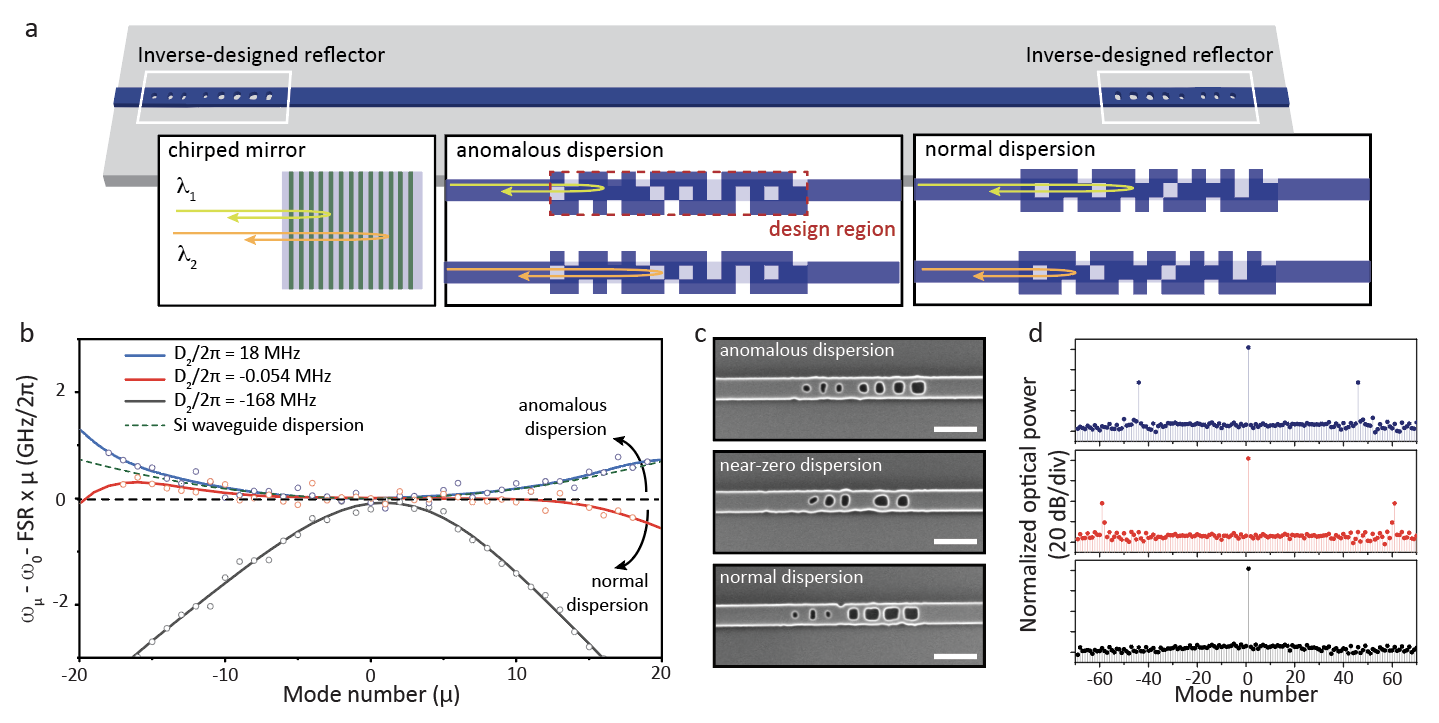}
%\captionsetup{format=plain,justification=RaggedRight}
\caption{\label{fig:Fig3}{\bf{Inverse designed dispersion engineered Fabry-Perot resonators}} (\textbf{a}) Schematic of inverse designed FP resonator and conceptual depiction of dispersion compensation from the inverse designed reflectors, analogous to chirped mirrors. (\textbf{b}) Measured integrated dispersion of three inverse designed FP resonators with cavity length of 200 $\mu m$ with inverse designed reflectors with dispersion targets. Blue, red, and black lines correspond to fit lines for the three FP resonators, showing anomalous, near-zero, and normal dispersion. (\textbf{c}) SEM images of the three different inverse designed reflectors used for the dispersion engineered resonators. (\textbf{d}) OPO simulation of the FP resonator with the experimental dispersion coefficients $D_{2}$ and $D_{3}$ derived from dispersion measurements of the three FP resonators with cavity length of 480 $\mu m$}
\end{figure*}

\noindent Due to their cylindrical symmetry, whispering gallery mode resonators possess degenerate resonant modes which propagate in the clockwise and counter-clockwise directions. By systematically coupling these two modes, we can control the positions of their resonances and engineer the desired dispersion profile. Here, we introduce a computational method to achieve this through inverse designed intracavity partially reflective elements (PREs)\cite{Yang:2020:NaturePhotonics}. 
%When there is intracavity reflection, the mode degeneracy is lifted and the counter directional propagating modes are hybridized proportional to the coupling strength dictated by the intracavity reflectance. 
%This phenomena is seen in many experiments with fabricated ring resonators where there is fabrication imperfection that leads to unwanted reflection in the resonator that hybridize the counter propagating resonance modes in the resonator. Here, we propose an intentional use of intracavity defect with the optimized reflection to be utilized in the microring resonator. 

\noindent When a PRE is introduced in the resonator, it couples the counter propagating resonant modes, lifting the mode degeneracy. The coupling strength, which is dictated by the reflectivity of the element, in turn determines the position of resonances of the resulting hybridized modes. More specifically, as we show in supplementary section I, the resonance splitting between the hybridized modes is given by $\Gamma = \theta/\pi \times \text{FSR}$, where FSR is the free-spectral range of the resonator and $\theta = \arctan\left( {|r|}/{\sqrt{1 - |r|^2}} \right)$, with $|r|$ being the reflectance of the PRE at the resonant frequency of the unhybridized modes. Consequently, a desired microring dispersion, specified as the spectral position of resonant modes, can be translated to a target frequency-dependent reflectance of the PRE, which can then be inverse designed. As a demonstration, we inverse design a PRE that has higher reflectance at higher frequency and lower reflectance at lower frequency. FIG 1a shows the scanning electron microscope (SEM) image of the inverse design PRE embedded in a microring resonator fabricated on a SOI stack.  FIG 1b shows the transmission spectrum of the microring resonator when excited through the coupled bus waveguide -- we indeed observe that the splitting between the hybridized microring resonator modes, and hence their resonant position, is dictated by the engineered reflectance spectrum of the PRE (FIG 1c). The measured integrated microring resonator dispersion ($D_{int}$) for the red-shifted and blue-shifted hybridized modes is depicted in FIG 1d --- we clearly observe a correlation between the extent of mode splitting and the reflectance spectrum.

\begin{figure*}[t!]
\centering
\includegraphics[width=\linewidth]{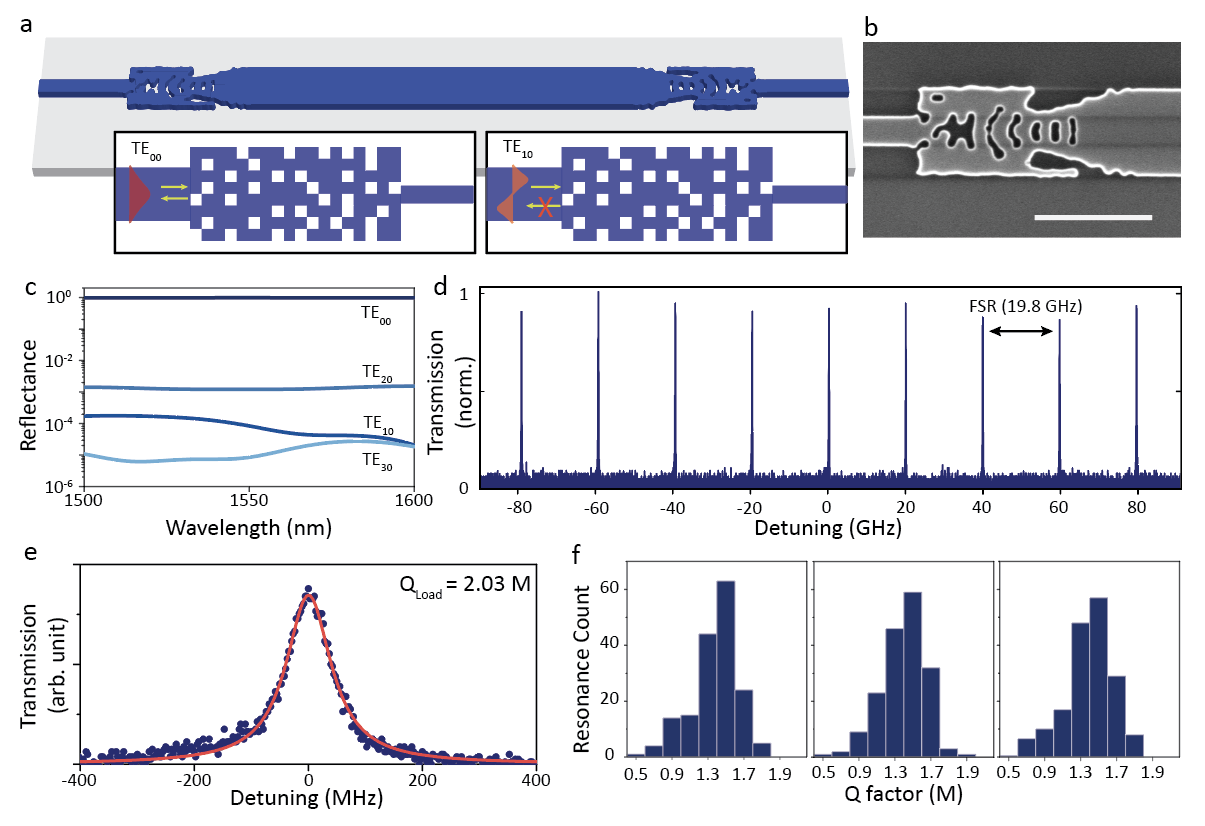}
%\captionsetup{singlelinecheck=no, justification = RaggedRight}
\caption{\label{fig:Fig4}{\bf{Inverse design of high Q FP resonator with single mode operation.}} \textbf{(a)} Schematic of the inverse designed FP resonator with inverse designed multimode reflectors allowing for a wide width waveguide for the resonator and single mode waveguide outputs. Conceptual picture showing the functionality of reflector, as it reflects fundamental mode well, and can couple out fundamental mode, while other higher order modes are not supported. \textbf{(b)} SEM image of the single mode inverse designed reflector with wider width waveguide (1.5 $\mu m$) for reflection of fundamental mode. Single mode waveguide of 0.5 $\mu m$ width is used on the other side to couple out the resonances from the resonator. The scale bar is 2 $\mu m$ \textbf{(c)} Simulated reflectance of the depicted inverse design reflector when excited with different TE modes, showing its selective reflection of the fundamental mode only. \textbf{(d)} Zoomed in transmission spectrum through the FP resonator, showing single mode operation separated by a FSR of 19.8 GHz. \textbf{(e)} Q-factor fit of a resonance of the inverse designed FP resonator, showing a loaded Q-factor of 2.03M. \textbf{(f)} Q factor statistics of resonance distribution from three different fabrication runs of the identical FP resonator design. }
\end{figure*}

\noindent A similar design technique can be applied to control the dispersion of on-chip fabry-perot (FP) resonators, which are composed of two reflectors with a straight waveguide section in the middle. This length of waveguide $L_D$, together with the group index $n_g$ of the propagating waveguide mode controls the FSR of the microresonator, as FSR = $c/2n_{g}L_{D}$. Traditionally, the dispersion of the FP resonator is controlled by engineering the waveguide cross-section and the length of the waveguide\cite{Papp:2019:ACSPhotonics}, or by employing chirped mirrors \cite{Chirped1994}. To achieve finer spectral control over the dispersion of the resonator, we can alternatively use inverse-design to engineer the frequency-dependent phase imparted by the mirrors at the ends of the FP resonator (FIG \ref{fig:Fig3}a). Such resonator with inverse designed reflectors is schematically shown in Fig. \ref{fig:Fig3}a. Notably, this concept of dispersion engineering using optimized reflectors with different dispersion targets is a generalization of chirped mirrors used in pulse engineering in lasers, where the distribution of multilayer coatings of thin films is optimized to provide engineered group delay dispersion \cite{Chirped1994}. As demonstrative examples, we use inverse designed reflectors to make the FP resonator dispersion anomalous, where group velocity dispersion is positive, or be normal, where group velocity dispersion is negative. In addition, the optimized reflectors can be very compact, with their width and length being only 0.5 $\mu m$ and 3.0 $\mu m$ respectively. The optimized dispersion engineered reflectors work by compensating phase upon the reflection. Notably, the inverse design technique finds the optimal structure that provides the desired wavelength dependent phase compensation in a compact footprint manner. Fig. \ref{fig:Fig3}b illustrates the experimental integrated dispersion profiles of the microresonators formed by three different inverse designed reflectors with different dispersion target, as shown in the SEM images of the reflectors in Fig. \ref{fig:Fig3}c. The cavity length used is 200 $\mu m$. We demonstrate tunable dispersion in equal length cavities ranging from anomalous to near-zero to normal, with $D_{2}$ values ranging from 18 MHz to -168 MHz. In addition, the length of the cavities can be varied,  achieving even more flexible dispersion tuning, as shown in FIG S2. To explore the potential implications of such dispersion engineered microresonators, we simulate optical parametric oscillation (OPO) based on the experimentally derived dispersion coefficients. Detailed implementation of the OPO simulation of a Fabry-Perot resonator with two photon absorption of silicon is shown in supplementary section II. Fig. \ref{fig:Fig3}d shows the result of OPO simulations with three different dispersion profiles derived from the inverse designed resonators with a 480 $\mu m$ cavity length, as shown in FIG S2. Importantly, when the dispersion of the microresonator is engineered to be normal, we do not see any OPO side band generation, as expected. When the dispersion of the microresonator is engineered to be anomalous at different strengths, the microresonators show OPOs generated at different resonance modes. \\

\noindent{\bf High Quality Factor Single Mode Resonator.}

\begin{figure*}[t!]
\centering
\includegraphics[width=\linewidth]{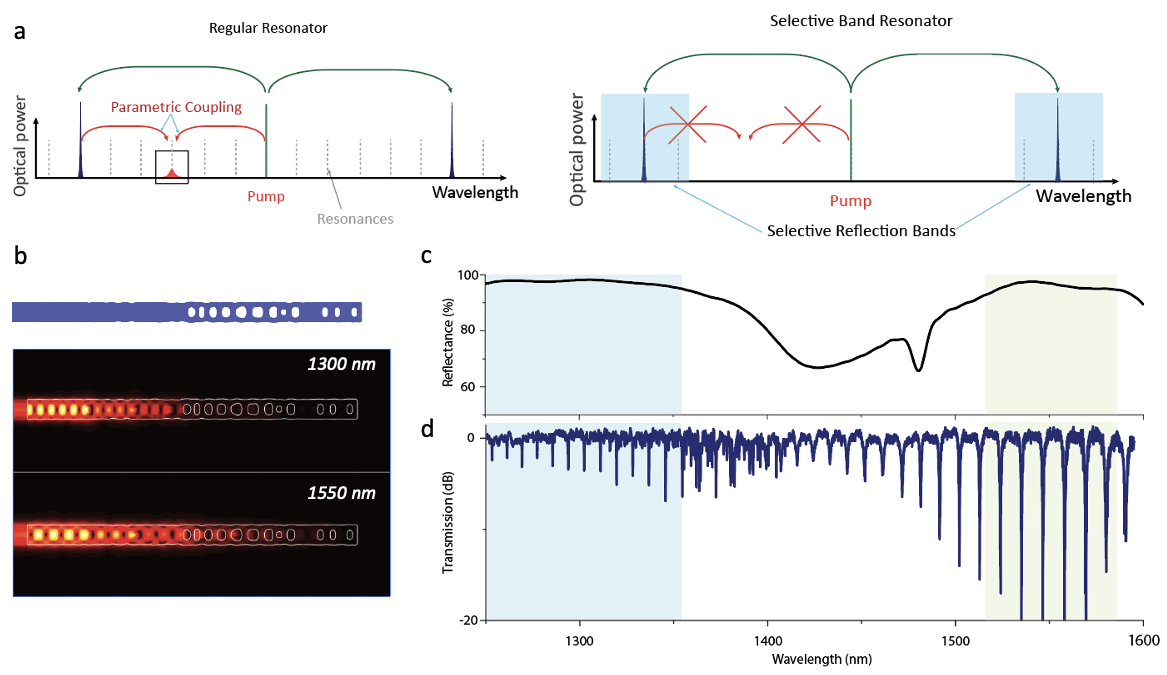}
%\captionsetup{format=plain,justification=RaggedRight}
\caption{\label{fig:Fig5}{\bf{Inverse design of selective wavelength band resonator}} (\textbf{a}) A graphical representation of parametric coupling in a typical resonator and in a selective band resonator, formed by selective reflection bands. (\textbf{b}) Schematic of utilized inverse designed selective wavelength reflector and its simulated electric field profile at the chosen wavelength of operation ($1300$ $nm$ and $1550$ $nm$).  (\textbf{c}) Simulated reflectance profile of the inverse designed selective reflection band reflector, showing high reflectance at O-band and C-band with low reflectance in other wavelength bands. (\textbf{d}) Measured transmission spectrum of the selective wavelength band resonator, showing sharp resonances in both O-band and C-band, while the resonances outside these bands are broad.}
\end{figure*}

\noindent To achieve high microresonator quality factors, the key is reducing the interaction of the optical fields with the microresonator surface, thereby reducing scattering losses due to surface roughness. As previously mentioned, this can often be achieved by using wide-width waveguides which tend to support fundamental waveguide modes that are well confined within the waveguide. However, wider waveguides often support multiple waveguide modes, which can couple to the fundamental mode and introduce loss.

\noindent To demonstrate how inverse design can be used to support this goal, we design a low loss FP resonator built out of a wide-width waveguide that supports only a single resonant mode. This is achieved by inverse designing reflectors that selectively reflect the fundamental mode of the wide waveguide and couple the transmitted field directly to single mode waveguides, suppressing higher-order modes in the process (FIG \ref{fig:Fig4}a). FIG \ref{fig:Fig4}b shows the SEM of the inverse designed reflector fabricated on a SOI stack, and FIG \ref{fig:Fig4}c shows its simulated reflectance when excited with different waveguide modes. We clearly see that the reflector only acts as a near-perfect mirror for the fundamental waveguide mode, while the remaining waveguide modes are allowed to transmit through and hence do not interfere with the fundamental mode operation. Furthermore, this functionality is achieved within a footprint of 3.5 $\mu m$ by 1.5 $\mu m$, which is significantly smaller than the conventional tapers that can be used to accomplish the same task. FIG \ref{fig:Fig4}d shows the transmission spectrum of the FP resonator and FIG \ref{fig:Fig4}e shows the zoomed-in spectrum of a single resonator mode --- we experimentally achieve a loaded Q-factor of 2.03 million at telecommunication wavelengths in the foundry-compatible SOI system with our design. Furthermore, the design is robust to fabrication variation, which is demonstrated in FIG \ref{fig:Fig4}f, where we measure the Q-factor of different fabrication runs of the same design and consistently achieve a high Q factor. \\

\noindent{\bf Selective Multiple Wavelength Band Operation.}

%In various applications, it is difficult to engineer selective wavelength band operations for the microresonators. For instance, microring resonators form resonances across all wavelength bands where there exists an optical mode for the resonator. While this is crucial for a broadband application such as octave spanning frequency comb generation \cite{Pfeiffer:17}, the existence of additional resonances may lead to parasitic coupling of the resonances to the undesired modes, as shown schematically in the top panel of Fig 5b. This parasitic coupling reduces the efficiency of the nonlinear wave mixing process such as second harmonic generation and OPO at a desired mode, limiting the realization of a high power on-chip light generation based on the nonlinear processes \cite{mckenna2021ultralowpower}. On the other hand, typical FP resonators only support single wavelength band, as the majority of mirror designs for the FP resonators rely on bragg grating mirror designs that are composed of a particular periodicity for a certain optical bandgap and hence supporting multiple wavelength bands for FP resonators have not been demonstrated before. 

\noindent The last design problem that we address is that of a microresonator with resonant frequencies in different frequency bands. One possible application of such a device would be to reduce pump leakage into undesired modes during a non-linear wave mixing process and decrease conversion efficiency from the mode competition process \cite{mckenna2021ultralowpower, stone2021conversion}. This is schematically depicted in FIG \ref{fig:Fig5}a, wherein we contrast between parametric down-conversion in a regular resonator, where coupling to many undesired resonant modes is possible, and in a band-selective resonator, where such couplings are absent. To achieve this with FP resonators, we can utilize photonic inverse design to optimize for high reflectance at chosen wavelength bands, and suppress the reflection at undesired wavelength bands. As a demonstration, we optimize a silicon reflector that has high reflectance at ($1300$ $nm$ and $1550$ $nm$) wavelength bands, which correspond to O-band and C-band, and reflectance suppression in the middle band ($1410$ $nm$). The resulting inverse-designed device, with a footprint of $8 \mu$m $\times$ 0.5 $\mu$m is shown in FIG \ref{fig:Fig5}b along with its simulated electric field profiles at the selected wavelengths. As shown in the electric field profiles, the single mirror reflects two well separated wavelengths ($1300$ $nm$ and $1550$ $nm$) effectively; its simulated reflectance spectrum is shown in FIG \ref{fig:Fig5}c. The schematic of the FP formed by the reflector is shown in FIG \ref{fig:FigS3}. To optically characterize the resonator, we side coupling the resonator to a waveguide-cavity directional coupler. We note that due to the difficulty in coupling equally to a resonator across distant wavelength bands \cite{Srinivasan:2019:NaturePhotonics} the resonances in O-band is weakly coupled, while the resonances in the C-band is close to critically coupled. This can be mediated by engineering coupler designs further \cite{Srinivasan:2019:NaturePhotonics}. FIG \ref{fig:Fig5}d depicts the measured transmission spectrum of the selective wavelength band microresonator. This resonator can achieve moderate loaded Q-factor in both of the selected bands (Q=1.5 $ \sim $  7K), while the suppressed wavelength bands show loaded quality factor as low as 600, which provides more than an order of magnitude difference between the selected and suppressed wavelength bands.  \\

\noindent{\bf Conclusion}

\noindent In summary, we demonstrate photonic inverse design approaches for on-chip microresonator designs that enable flexible engineering of important figure of merits such as dispersion, high Q-factor operation and selective wavelength band operation. As the proposed techniques can be combined together to engineer for multiple desirable metrics at the same time, we foresee the combined optimized techniques achieving nearly arbitrary resonator designs. Furthermore, given this is the demonstration of a general approach not limited to the SOI system used here as a model platform, it can easily extend to other material platforms including those with transparency in the visible spectrum and those with enhanced nonlinear properties like second order nonlinearity. For instance, selective band operation can be extended to platforms such as lithium niobate\cite{Loncar:2018:Nature} and silicon carbide\cite{Lukin:2020:NaturePhotonics}, where we can target f-2f nonlinear operations in visible and telecom wavelength bands. Even in silicon, achieving a near-zero dispersion resonator opens up the possibility of resonator enhanced EOM combs \cite{Afltouni:2020:CLEO,Loncar:2019:Nature} and mid-IR nonlinear wave mixing\cite{Bell:Natpho} in commercial foundries, which is critical for on-chip communications and spectroscopy.

\bibliography{Reference}

\clearpage

\noindent{\bf Methods}\\

\noindent\textbf{Photonic Inverse Design} \\
Stanford Photonics Inverse DesignSoftware (SPINS) was utilized to optimize the various reflectors, and intracavity element shown through out the work. SPINS can provide design of various on-chip elements as previously demonstrated. \\

\noindent\textbf{Data availability} The data that support the plots within this paper and other findings of this study are available from the corresponding author upon reasonable request.\\

\noindent\textbf{Code availability} An open source of the photonic optimization software used in this paper is available at https://github.com/stanfordnqp/spins-b.\\

\medskip

\noindent\textbf{Acknowledgment}
\noindent  We acknowledge insightful discussion with F.Aflatouni, S.Papp, and K.Srinivasan. This work is funded by the DARPA under the PIPES and LUMOS program.  We thank G.Keeler and the programme management teams for discussions throughout the project. G. H. A. acknowledges support from STMicroelectronics Stanford Graduate Fellowship (SGF) and Kwanjeong Educational Foundation. R. T. acknowledges Max Planck Harvard research center for quantum optics (MPHQ) postdoctoral fellowship. A. W. acknowledges the Herb and Jane Dwight Stanford Graduate Fellowship. J. S. acknowledges support from Cisco Systems Stanford Graduate Fellowship (SGF)\\

\noindent\textbf{Competing interests.}
The authors declare they have no competing financial interests.

\clearpage

\clearpage
\onecolumngrid 

\appendix 
\renewcommand{\thefigure}{S\arabic{figure}}
\renewcommand{\thesection}{\Roman{section}}
\setcounter{figure}{0} 
\section* {Supplementary Information to\\Augmenting On-Chip Microresonator through Photonic Inverse Design}
\vspace{-0.15 in}
\noindent Geun Ho Ahn$^{1}$, Ki Youl Yang$^{1}$, Rahul Trivedi$^{1,2}$, Alexander D. White$^{1}$, Logan Su$^{1}$, Jinhie Skarda$^{1}$, Jelena Vu\v{c}kovi\'{c}$^{1,\dagger}$\\
\vspace{0.1 in}\\
\noindent
$^1$E. L. Ginzton Laboratory, Stanford University, Stanford, CA 94305, USA.\\
$^2$Max-Planck-Institute for Quantum Optics, Hans-Kopfermann-Str. 1, 85748 Garching, Germany\\
$\dagger$Correspondence and requests for materials should be addressed to jela@stanford.edu.\\

\section*{Section I: Modelling Partially Reflecting Element in microring resonators}
  \noindent \emph{Coupled mode equations for intracavity element}: Here we develop coupled mode theory for a general intracavity element in a whispering galley resonator. As shown in the Fig S1, the counter directional propagating modes (clockwise and counter clockwise propagating modes) $b(\omega)$ and $a(\omega)$ can be formulated as:

  \begin{subequations}
  \begin{align}
  & a_{in}(\omega) = a_{out}(\omega) \times e^{i\phi(\omega)} \\
  & b_{in}(\omega) = b_{out}(\omega) \times e^{i\phi(\omega)}
  \end{align}
 
  \end{subequations}

\noindent The counter direction propagating modes can be coupled through the intracavity element with its optical characteristics  $|t|^2 + |r|^2 = 1$ with following coupled mode relation:

 \begin{subequations}
  \begin{align}
  & a_{out}(\omega) = t \times a_{in}(\omega) + r \times b_{in}(\omega) \\
  & b_{out}(\omega) = -r^* \times a_{in}(\omega) + t \times b_{in}(\omega)
  \end{align}
 
  \end{subequations}

\noindent Solving for the eigenvalue which is the coupling strength strength leads to the coupling $\theta = arctan(\frac{|r|}{|t|})$. \\

\noindent The resonance conditions with and without the PRE are

  \begin{align}
  \phi_{\pm}(\omega) = 2n\pi \pm \theta
  \end{align}

  \begin{align}
  \phi(\omega_n) = 2n\pi
  \end{align}

\noindent We use taylor expansion of $\phi_{\pm}(\omega)$ to obtain:

  \begin{align}
  \phi_{\pm}(\omega) = \phi_{\pm}(\omega_n) + (\omega - \omega_n)\phi'(\omega_n) = 2n\pi \pm \theta
  \end{align}

  \begin{align}
   (\omega_{\pm} - \omega_n) =\pm \frac{\theta}{\phi'(\omega_n)}
  \end{align}

where

  \begin{align}
   \phi'(\omega_n) = \frac{2\pi}{FSR} 
  \end{align}

Hence the final resonance splitting ($\Gamma$) is

  \begin{align}
    \Gamma = \omega_+ - \omega_- = \frac{arctan\left(\frac{|r|}{|t|}\right)}{\pi} \times FSR
  \end{align}

\begin{figure*}[t!]
\includegraphics[width=0.3 \linewidth]{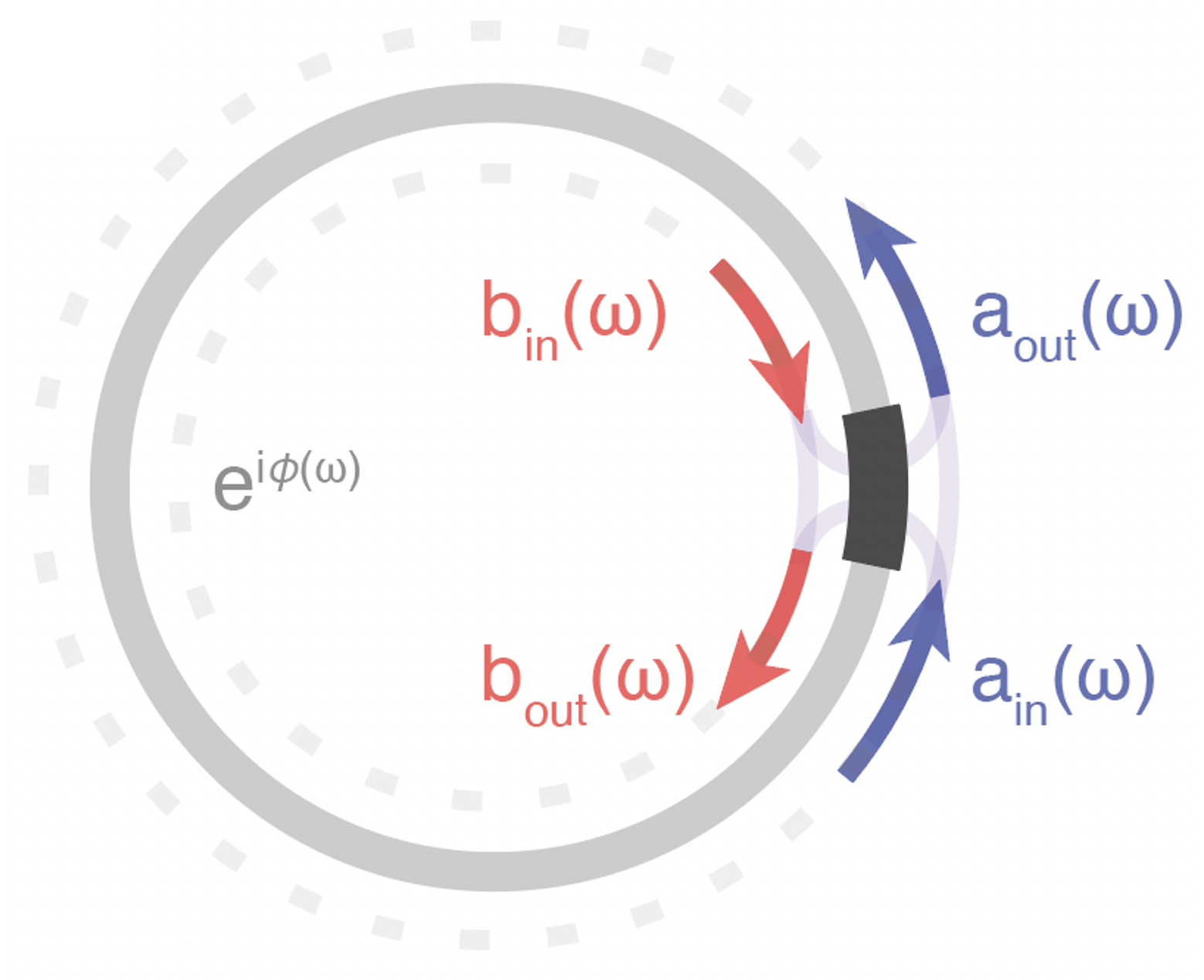}
\caption{\label{fig:FigS1}{\bf{Schematic of coupled mode theory for an intracavity element in whispering gallery mode resonator}}}
\end{figure*}

\clearpage
\begin{figure*}[t!]
\includegraphics[width=\linewidth]{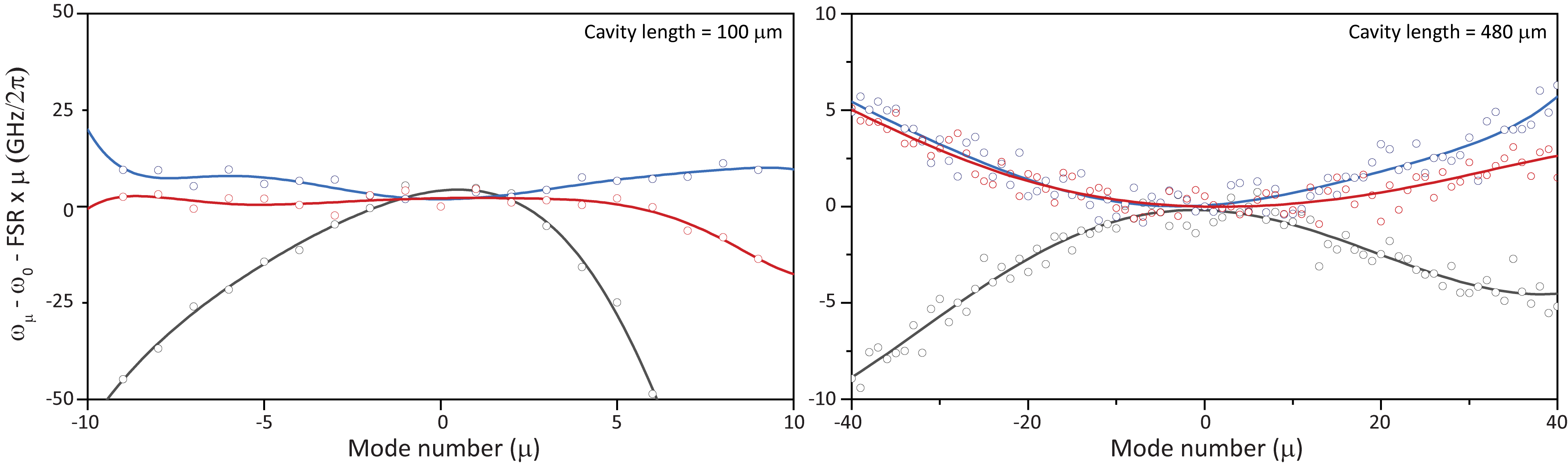}
\caption{\label{fig:FigS2}{\bf{Measured integrated dispersion of three inverse designed FP resonators with cavity length of 100 $\mu m$ and 480 $\mu m$}}}
\end{figure*}
\clearpage
\begin{figure*}[t!]
\includegraphics[width=\linewidth]{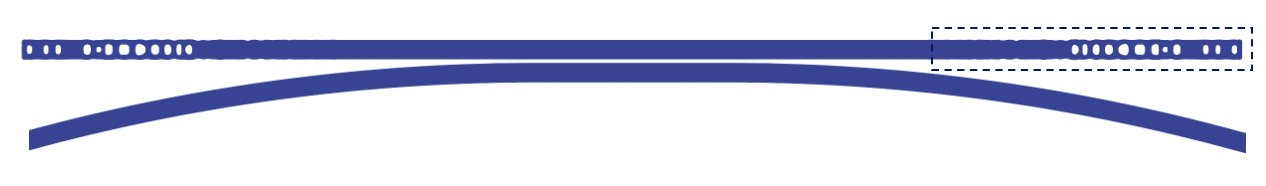}
\caption{\label{fig:FigS3}{\bf{Schematic of side coupled inverse designed FP resonator with selective reflection bands. This FP is coupled through a directional coupler due to the flexibility in tuning of coupling strength by tuning the gap size instead of targeting a particular through coupling strength. The gap size is chosen to be 100 $nm$. }}}
\end{figure*}
\clearpage
\section*{Section II: Modelling TPA in optical parametric oscillations simulations}
  \noindent \emph{Coupled mode equations for general cavities}: Here we develop coupled mode theory for a general resonator in the presence of TPA. We assume that the resonator is described by modes $\textbf{E}_\mu(\textbf{x})$ and resonant frequencies $\omega_\mu$ which satisfy:
  \begin{align}
  \nabla \times \nabla \times \textbf{E}_\mu(\textbf{x}) = \mu_0 \varepsilon_0 \omega_\mu^2 n^2(\textbf{x}) \textbf{E}_\mu(\textbf{x})
  \end{align}
  Furthermore, in the limit of a lossless cavity, the modes satisfy the following orthonormality relationship:
  \begin{align}
  \int_\Omega n^2(\textbf{x})\textbf{E}_\mu^*(\textbf{x})\cdot \textbf{E}_\nu(\textbf{x})d^3\textbf{x} = \delta_{\mu, \nu}
  \end{align}
where $\Omega$ denotes the cavity region. The starting point for developing the coupled mode theory is the Maxwell's equations for the electric field:
  \begin{align}
  \nabla \times\nabla \times \textbf{E}(\textbf{x}, t) + \mu_0 \varepsilon_0 n^2(\textbf{x}) \frac{\partial^2}{\partial t^2} \textbf{E}(\textbf{x}, t) = -\mu_0 \frac{\partial}{\partial t}\textbf{J}_{\text{TPA}}(\textbf{x}, t) - \mu_0 \frac{\partial^2}{\partial t^2} \textbf{P}_{\text{NL}}(\textbf{x}, t)
  \end{align}
  As a model for TPA, we setup a current that varies nonlinearly with $\textbf{E}(\textbf{x}, t)$:
  \begin{align}\label{eq:tpa_curr}
  \textbf{J}_{\text{TPA}}(\textbf{x}, t) = \begin{cases} 
  \sigma_{\text{TPA}} \big(\textbf{E}(\textbf{x}, t) \cdot \textbf{E}(\textbf{x}, t)\big)\textbf{E}(\textbf{x}, t) & \text{if } \textbf{x} \in \Omega_{\text{NL}} \\
  0 & \text{otherwise}
  \end{cases}
  \end{align}
 where $\Omega_{\text{NL}}$ is the region inside the cavity volume where the nonlinear material is. Similarly, $\textbf{P}_{\text{NL}}(\textbf{x}, t)$ has the third order nonlinear form --- we assume an isotropic polarization:
  \begin{align}
  \textbf{P}_{\text{NL}}(\textbf{x}, t)=
  \begin{cases}
  \varepsilon_0 \chi \big(\textbf{E}(\textbf{x}, t) \cdot \textbf{E}(\textbf{x}, t)\big)\textbf{E}(\textbf{x}, t) & \text{if } \textbf{x} \in \Omega_{\text{NL}} \\
 0 & \text{otherwise} 
  \end{cases} 
  \end{align}
  Next, we write the electric field as a superposition of cavity modes:
  \begin{align}
  \textbf{E}(\textbf{x}, t) = \underbrace{\bigg(\frac{\hbar \omega_0}{2\varepsilon_0}\bigg)^{1/2}\sum_\mu A_\mu(t) e^{-i\omega_\mu t} \textbf{E}_\mu(\textbf{x})}_{\tilde{\textbf{E}}(\textbf{x}, t)} + \text{c.c.}
  \end{align}
  where $\tilde{\textbf{E}}(\textbf{x}, t)$ has only positive frequencies. We note that $A_\mu(t)$ are dimensionless in this expression and $|A_\mu(t)|^2$ is the number of photons in the $\mu^{\text{th}}$ cavity mode at time $t$. Furthermore, we also assume that all the frequencies in this term are very close to a central frequency $\omega_0$. With this form of the field, the contribution of the following nonlinear expression in the electric field to frequencies close to $\omega_0$ can be easily calculated:
  \begin{align}
  \big(\textbf{E}(\textbf{x}, t) \cdot \textbf{E}(\textbf{x}, t)\big)\textbf{E}(\textbf{x}, t) \approx 2 | \tilde{\textbf{E}}(\textbf{x}, t)|^2 \tilde{\textbf{E}}(\textbf{x}, t)
  \end{align}
  Here we have only retained the intensity dependent refractive index contributions, and the parametric couplings in the nonlinear term, and ignored harmonic generation. Defining
  \begin{align}
  \tilde{\textbf{J}}_{\text{TPA}}(\textbf{x}, t) = \begin{cases} 
  2\sigma_{\text{TPA}} |\tilde{\textbf{E}}(\textbf{x}, t)|^2\tilde{\textbf{E}}(\textbf{x}, t) & \text{if } \textbf{x} \in \Omega_{\text{NL}} \\
  0 & \text{otherwise} 
  \end{cases} \ \text{and} \ 
  \tilde{\textbf{P}}_{\text{NL}} = \begin{cases}
   2\varepsilon_0 \chi |\tilde{\textbf{E}}(\textbf{x}, t)|^2 \tilde{\textbf{E}}(\textbf{x}, t) & \text{if } \textbf{x} \in \Omega_{\text{NL}} \\
   0 & \text{otherwise}
   \end{cases}
  \end{align}
  we obtain:
  \begin{align}\label{eq:one_sided_eq}
   \nabla \times\nabla \times \tilde{\textbf{E}}(\textbf{x}, t) + \mu_0 \varepsilon_0 n^2(\textbf{x}) \frac{\partial^2}{\partial t^2} \tilde{\textbf{E}}(\textbf{x}, t) = i\omega_0\mu_0 \tilde{\textbf{J}}_{\text{TPA}}(\textbf{x}, t) + \mu_0\omega_0^2  \tilde{\textbf{P}}_{\text{NL}}(\textbf{x}, t)
\end{align}
Before proceeding further, we point out relationships between the model parameters $\sigma_{\text{TPA}}$ and $\chi$ to the experimentally measured $\beta_{\text{TPA}}$ (change in absorption coefficient per unit intensity) and $n_2$ (change in refractive index per unit intensity). These parameters can only be defined rigorously for plane-waves --- supposes we consider a problem where $n(\textbf{x})$ is uniform in space with value $n_0$ and consider solutions of the form $\tilde{E}(\textbf{x}, t) = \hat{z}\mathcal{E}_0(x) e^{ -i \omega t}$ to the above equation then Eq.~\ref{eq:one_sided_eq} reduces to:
\begin{align}\label{eq:1d_model}
\frac{d^2 \mathcal{E}(x)}{dx^2} + \frac{\omega_0^2}{c^2}\bigg[n_0^2 + \bigg(2\chi  + \frac{2i \sigma_{\text{TPA}}}{\omega_0 \varepsilon_0}\bigg)|\mathcal{E}(x)|^2\bigg]  \mathcal{E}(x) = 0
\end{align}
Furthermore, assuming $\mathcal{E}(x)$ to be very close to a plane-wave solution $e^{ik_0 x}$ where $k_0 = \omega n_0 / c$, we may calculate the intensity in the electromagnetic field $I$ via:
\begin{align}
I(x) = 2\epsilon_0 cn_0 |\mathcal{E}(x)|^2
\end{align}
Now, from Eq.~\ref{eq:1d_model} we may conclude that the local refractive index $n(x)$ depends on the intensity $I(x)$ via:
\begin{align}
n(x) \approx n_0 + \frac{\chi |\mathcal{E}(x)|^2}{n_0} = n_0 + \frac{\chi}{2\varepsilon_0 c n_0^2}I(x)
\end{align}
which yields $\chi = 2 n_0^2 c \varepsilon_0 n_2$. Similarly, from the imaginary part of the local refractive index, we can calculate the intensity dependent absorption coefficient (dimensions of 1 / length):
\begin{align}
\alpha(x) = \frac{2\sigma_{\text{TPA}}}{c\varepsilon_0 n_0} |\mathcal{E}(x)|^2 = \frac{\sigma_{\text{TPA}}}{c^2\varepsilon_0^2 n_0^2} I(x)
\end{align}
which yields $\sigma_{\text{TPA}} = c^2 \varepsilon_0^2 n_0^2 \beta_{\text{TPA}}$.
 Note that:
 \begin{align}
 \nabla \times \nabla \times \tilde{\textbf{E}}(\textbf{x}, t) + \mu_0 \varepsilon_0 n^2(\textbf{x}) \frac{\partial^2}{\partial t^2}\textbf{E}(\textbf{x}, t) \approx -2i\mu_0 \varepsilon_0 \omega_0 n^2(\textbf{x})\bigg(\frac{\hbar \omega_0}{2\varepsilon_0}\bigg)^{1/2} \sum_\mu  \frac{dA_\mu(t)}{dt} e^{-i\omega_\mu t} \textbf{E}_\mu(\textbf{x})
 \end{align}
 and thus
 \begin{align}\label{eq:first_eq}
 \frac{dA_\mu(t)}{dt} = \bigg(\frac{2\varepsilon_0}{\hbar \omega_0}\bigg)^{1/2}\frac{\omega_0 (i\varepsilon_0 \omega_0 \chi - \sigma_{\text{TPA}})e^{i\omega_\mu t}}{\omega_\mu \varepsilon_0} \int \big|\tilde{\textbf{E}}(\textbf{x}, t)\big|^2 \textbf{E}_\mu^*(\textbf{x})\cdot \tilde{\textbf{E}}(\textbf{x}, t)d^3\textbf{x}
 \end{align}
 We now consider the evaluation of the term in the integral: By a straightforward substitution of the expression for $\tilde{\textbf{E}}(\textbf{x}, t)$, we obtain:
 \begin{align}
 \int \big|\tilde{\textbf{E}}(\textbf{x}, t)\big|^2 \textbf{E}_\mu^*(\textbf{x})\cdot \tilde{\textbf{E}}(\textbf{x}) d^3\textbf{x} =\bigg(\frac{\hbar \omega_0}{2\varepsilon_0}\bigg)^{3/2} \frac{1}{n_0^4 V_0}\sum_{\nu, \alpha, \beta} \Lambda_{\mu, \nu, \alpha, \beta} A_\alpha^*(t) A_\nu(t) A_\beta(t) e^{i(\omega_\alpha - \omega_\nu - \omega_\beta) t}
 \end{align}
 where
 \begin{align}
 \Lambda_{\mu. \nu, \alpha, \beta} = \frac{\int_{\Omega_{\text{NL}}}\big[\textbf{E}_\alpha^*(\textbf{x})\cdot \textbf{E}_\beta(\textbf{x})\big] \big[\textbf{E}_\mu^*(\textbf{x})\cdot \textbf{E}_\nu(\textbf{x})\big] d^3\textbf{x}}{\int_{\Omega_{\text{NL}}}  |\textbf{E}_0^*(\textbf{x})\cdot \textbf{E}_0(\textbf{x})|^2 d^3 \textbf{x}}
 \end{align}
 and
 \begin{align}
 V_0 = \frac{1}{n_0^4}\bigg[{\int_{\Omega_{\text{NL}}}  |\textbf{E}_0^*(\textbf{x})\cdot \textbf{E}_0(\textbf{x})|^2d^3\textbf{x}}\bigg]^{-1} = \frac{1}{n_0^4} \bigg[\frac{\big(\int_\Omega n^2(\textbf{x}) \textbf{E}_0^*(\textbf{x})\cdot \textbf{E}_0(\textbf{x})d^3 \textbf{x}\big)^2}{\int_{\Omega_{\text{NL}}} |\textbf{E}_0^*(\textbf{x})\cdot \textbf{E}_0(\textbf{x})|^2d^3\textbf{x}} \bigg]
 \end{align}
Here $n_0$ is the refractive inside the cavity. We note that for tightly confined modes, $n^2(\textbf{x}) \approx n_0^2$ inside the integrals in the above expressions, and we obtain:
\begin{align}
\Lambda_{\mu, \nu, \alpha, \beta} \approx \frac{\int  \big[\textbf{E}_\alpha^*(\textbf{x})\cdot \textbf{E}_\beta(\textbf{x})\big] \big[\textbf{E}_\mu^*(\textbf{x})\cdot \textbf{E}_\nu(\textbf{x})\big] d^3\textbf{x}}{\int |\textbf{E}_0^*(\textbf{x})\cdot \textbf{E}_0(\textbf{x})|^2 d^3 \textbf{x}} \ \text{and} \ V_0 \approx {\frac{\big(\int \textbf{E}_0^*(\textbf{x})\cdot \textbf{E}(\textbf{x}) d^3\textbf{x}\big)^2}{\int |\textbf{E}_0^*(\textbf{x})\cdot \textbf{E}_0(\textbf{x})|^2d^3\textbf{x}}}
\end{align}
 Consequently, we obtain a nonlinear ODE for $A_\mu(t)$:
 \begin{align}
 \frac{dA_\mu(t)}{dt} = \frac{\hbar \omega_0(i\varepsilon_0 \omega_0 \chi - \sigma_{\text{TPA}})}{2n_0^4 \varepsilon_0^2 V_0} \sum_{\nu, \alpha, \beta} \Lambda_{\mu, \nu, \alpha, \beta} A_\alpha^*(t) A_\nu(t) A_\beta(t) e^{i(\omega_\alpha + \omega_\mu - \omega_\nu -\omega_\beta)t}
 \end{align}
 Next, we introduce the parametric coupling strength $g_0$ (with dimensions of 1 / time) as:
 \begin{align}
 g_0 = \frac{\hbar \omega_0^2 \chi}{2n_0^4 \varepsilon_0 V_0} = \frac{\hbar \omega_0^2 n_2 c}{n_0^2 V_0}
 \end{align}
 as well as the dimensionless parametric loss parameter $\xi$:
 \begin{align}
 \xi = \frac{\sigma_{\text{TPA}}}{\varepsilon_0 \omega_0 \chi} = \frac{\beta_{\text{TPA}} c}{2 \omega_0 n_2}
 \end{align}
 with which:
 \begin{align}
 \frac{dA_\mu(t)}{dt} = ig_0 ( 1 + i\xi) \sum_{\nu, \alpha, \beta} \Lambda_{\mu, \nu, \alpha, \beta} A_\alpha^*(t) A_\nu(t) A_\beta(t) e^{i(\omega_\alpha + \omega_\mu - \omega_\nu -\omega_\beta)t}
 \end{align}
 Note that uptil now, we have ignored any losses in the cavity mode except for the two-photon absorption loss, and have treated the linear cavity modes to be lossless. An important consideration while studying losses in silicon based devices is that the material absorption depends on the number of carriers in the structure, which can become large at high field intensities due to carrier generation based on two-photon absorption. Thus the decay rate of each cavity mode is written as $\kappa_\mu + \gamma_\mu N(t)$, where $N(t)$ is the number of carriers per unit volume inside the material:
 \begin{align}\label{eq:main_dynamics}
 \boxed{\frac{dA_\mu(t)}{dt} = -\frac{\kappa_\mu + \gamma_\mu N(t)}{2} A_\mu(t) +  ig_0 ( 1 + i\xi) \sum_{\nu, \alpha, \beta} \Lambda_{\mu, \nu, \alpha, \beta} A_\alpha^*(t) A_\nu(t) A_\beta(t) e^{i(\omega_\alpha + \omega_\mu - \omega_\nu -\omega_\beta)t}}
 \end{align}
A simple estimate of $\gamma_\mu$ can be made from Maxwell's equations by (i) ignoring the spatial distribution of the generated carriers and (ii) Assuming that the relaxation time-scale of the generated carriers is much faster than the dynamics of the cavity mode. Under these assumptions, the carriers simply introduce a conductivity $\sigma = N e^2 \tau / m_{\text{eff}}( 1 + \omega_0^2 \tau^2)$. This conductivity adds a current $\sigma \textbf{E}(\textbf{x}, t)$ to Maxwell's equations above, which introduces the following decay term in the RHS of equation for $dA_\mu(t) / dt$ (Eq.~\ref{eq:first_eq}) from which we obtain the following expression for $\gamma_\mu$:
\begin{align}
\gamma_\mu = \frac{\omega_0}{\omega_\mu} \bigg[\frac{e^2 \tau}{m_{\text{eff}}(1 + \omega_0^2 \tau^2)}\bigg] \int_{\text{Si}} |\textbf{E}_\mu(\textbf{x})|^2 d^3\textbf{x}
\end{align}
Next, we consider the dynamics of carrier generation due to two-photon absorption. The number of carriers generated per unit time, $F(t)$ can be estimated by calculating the total two-photon absorbed power and dividing it by $2\hbar \omega_0$ (energy of two photons being absorbed). Using the effective current in Eq.~\ref{eq:tpa_curr}:
\begin{align}
F(t) = \frac{\sigma_{\text{TPA}}}{2\hbar \omega_0} \bigg \langle \int (\textbf{E}(\textbf{x}, t) \cdot \textbf{E}(\textbf{x}, t))^2 d^3 \textbf{x}\bigg\rangle_{\text{opt. cycle}}
\end{align}
Here, we are assuming that the dynamics of the carrier generation is comparable to the dynamics of the cavity mode, but still much slower than an optical cycle. Thus, we perform an averaging over the optical cycle. Performing this averaging, we obtain:
\begin{align}
F(t) = \frac{4\sigma_{\text{TPA}}}{2\hbar \omega_0}  \int (\tilde{\textbf{E}}^*(\textbf{x}, t) \cdot \tilde{\textbf{E}}(\textbf{x}, t) )^2 d^3 \textbf{x}
\end{align}
Using the expansion for $\tilde{\textbf{E}}(\textbf{x}, t)$ in terms of the cavity mode amplitude and assuming the cavity mode to be tightly confined, we obtain:
\begin{align}
F(t) = \frac{\sigma_{\text{TPA}} \hbar \omega_0}{2\varepsilon_0^2 n_0^4 V_0} \sum_{\mu, \nu, \alpha, \beta} \Lambda_{\mu, \nu, \alpha, \beta} A_\mu^*(t) A_\alpha^*(t) A_\nu (t) A_\beta(t) e^{i(\omega_\mu + \omega_\alpha - \omega_\mu -\omega_nu)t}
\end{align}
We can then introduce the following differential equation for the carrier which accounts for an exponential decay of the carriers of the order of the dwell time $\tau_d$ of the carrier inside the material:
\begin{align}\label{eq:carrier_dynamics}
\boxed{\frac{dN(t)}{dt} = -\frac{N(t)}{\tau_d} +  g_0 \xi \sum_{\mu, \nu, \alpha, \beta} \Lambda_{\mu, \nu, \alpha, \beta} A_\mu^*(t) A_\alpha^*(t) A_\nu (t) A_\beta(t) e^{i(\omega_\mu + \omega_\alpha - \omega_\nu -\omega_\beta)t}}
\end{align}\\ \ \\
\noindent \emph{Fabry Perot resonator}: Several simplifications to the above coupled mode equations happens for Fabry perot resonators by accounting for phase-matching conditions between different resonant modes. These simplifications are important since they make the dynamics easier to simulate. Consider. Fabry-perot mode formed by mirrors at $x = 0$ and $x = L$ with mirros that impart a frequency dependent phase-shift $\phi(\omega)$. The resonator mode can be expressed as:
\begin{align}
\textbf{E}_\mu(\textbf{x}) = \textbf{E}^{(+)}_\mu(\rho) e^{i\beta_\mu x} + \textbf{E}^{(-)}_\mu(\rho) e^{-i\beta_\mu x} e^{-i\phi_\mu}
\end{align}
where $\textbf{E}_\mu^{(\pm)}(\rho), \beta_\mu, \phi_\mu$ are the waveguide modes, propagation constant and mirror phase evaluated at $\omega = \omega_\mu$ and $\rho \equiv (x, y)$ is the transverse coordinate. Furthermore, $\textbf{E}_\mu^{(-)} = \big[\textbf{E}_\mu^{(+)}\big]^*$.  We can use this form of the modes to evaluate $\Lambda_{\mu, \nu, \alpha, \beta}$ defined above. For this calculation, we will only keep terms in the evaluation $\Lambda_{\mu, \nu, \alpha, \beta}$ which are phase matched. Mathematically, phase-matching is a consequence of approximating the following integral in the limit of long length $L$:
\begin{align}
\int_0^L e^{i\Delta \beta x} dx = \frac{2 e^{i\Delta \beta L / 2} \sin \Delta \beta L / 2}{\Delta \beta} \approx \begin{cases} L & \text{if } \Delta \beta = 0 \\
0 & \text{otherwise}
\end{cases}
\end{align}
Furthermore, we also assume that $\beta_\mu$ and $\phi_\mu$ can be approximated well by a linear function of $\mu$ as far as phase-matching is concerned. This essentially means that the length of the resonator $L$ is much larger than the inverse of the linear contribution to $\beta_\mu$, while being smaller than the non-linear contributions. This immediately allows us to write the following approximations to various integrals arising in the calculation of $\Lambda_{\mu, \nu, \alpha, \beta}$:
\begin{subequations}
\begin{align}
&\int_0^L e^{\pm i(\beta_\mu + \beta_\nu + \beta_\alpha + \beta_\beta)x} dx \approx 0 \\
&\int_0^L e^{\pm i (\beta_\mu + \beta_\nu + \beta_\alpha - \beta_\beta)x} dx \approx 0 \\
&\int_0^L e^{i(\beta_\mu + \beta_\nu - \beta_\alpha - \beta_\beta) x} dx \approx L \delta_{\mu + \nu, \alpha + \beta}
\end{align}
\end{subequations}
A simple calculation then yields:
\begin{align}
\Lambda_{\mu, \nu, \alpha, \beta} = \text{M}_{\mu, \nu, \alpha, \beta} \delta_{\mu + \alpha, \nu + \beta} + \text{N}_{\mu, \nu, \alpha, \beta} \delta_{\mu + \beta, \alpha + \nu} + \text{P}_{\mu, \nu, \alpha, \beta} \delta_{\mu + \nu, \alpha + \beta}
\end{align}
where
\begin{align}
&\text{M}_{\mu, \nu, \alpha, \beta} = L\frac{\int  \big[(\textbf{E}_\mu^{(+)*}(\rho)\cdot \textbf{E}_\nu^{(+)}(\rho))(\textbf{E}_\alpha^{(+)*}(\rho)\cdot \textbf{E}^{(+)}_\beta(\rho)) + \text{c.c.}\big]d^2\rho}{\int |\textbf{E}_0^*(\textbf{x})\cdot \textbf{E}_0(\textbf{x})|^2d^3\textbf{x}} \nonumber \\
&\text{N}_{\mu, \nu, \alpha, \beta} = L\frac{\int  \big[e^{i(\phi_\alpha - \phi_\beta)}(\textbf{E}_\mu^{(+)*}(\rho)\cdot \textbf{E}_\nu^{(+)}(\rho)) (\textbf{E}_\alpha^{(-)*}(\rho) \cdot \textbf{E}^{(-)}_\beta(\rho)) + \text{c.c.}\big]d^2 \rho}{\int |\textbf{E}_0^*(\textbf{x})\cdot \textbf{E}_0(\textbf{x})|^2d^3\textbf{x}}\nonumber \\
&\text{P}_{\mu, \nu, \alpha, \beta} = L\frac{\int  \big[e^{i(\phi_\alpha - \phi_\nu)}(\textbf{E}_\mu^{(+)*}(\rho)\cdot \textbf{E}_\nu^{(-)}(\rho)) (\textbf{E}_\alpha^{(-)*}(\rho)\cdot \textbf{E}_\beta^{(+)}(\rho)) + \text{c.c.}\big]d^2\rho}{\int |\textbf{E}_0^*(\textbf{x})\cdot \textbf{E}_0(\textbf{x})|^2d^3\textbf{x}}
\end{align}
Another standard approximation that is needed to obtain an LLE like form is to ignore dependence of $\text{M}_{\mu, \nu, \alpha, \beta}, \text{N}_{\mu, \nu, \alpha, \beta}$ and $\text{P}_{\mu, \nu, \alpha, \beta} $ on the mode numbers and approximate them by their value at the central mode. Doing so, we obtain that:
\begin{align}
\text{M}_{\mu, \nu, \alpha, \beta} \approx \text{N}_{\mu, \nu, \alpha, \beta} \approx \text{P}_{\mu, \nu, \alpha, \beta} \approx \Lambda_0 = 2L \frac{\int (\textbf{E}^{(+)*}_0(\rho)\cdot\textbf{E}_0^{(+)}(\rho))^2d^2 \rho}{\int |\textbf{E}_0^*(\textbf{x})\cdot \textbf{E}_0(\textbf{x})|^2d^3\textbf{x}}
\end{align}
Using the form of the Fabry-perot modes, it can be seen that:
\begin{align}
\int |\textbf{E}_0^*(\textbf{x})\cdot \textbf{E}_0(\textbf{x})|^2 d^3 \textbf{x} = 2L \bigg(1 - \frac{\sin \phi_0}{\beta_0 L}\bigg) \int \textbf{E}_0^{(+)*}(\rho)\cdot \textbf{E}^{(+)}_0(\rho) d^2 \rho 
\end{align}
From this it follows that for long resonators ($\beta_0 L \gg 1$), $\Lambda_0 \approx 1$.
Consider now the nonlinear term in Eq.~\ref{eq:main_dynamics} --- we now make an assumption on the resonance frequencies $\omega_\mu$: we assume that it is approximately linear with $\mu$:
\begin{align}
\omega_\mu = \omega_0 + D_1 \mu + \delta_\mu
\end{align}
with $\delta_\mu \ll D_1$. We assume that $D_1$ is much larger than the rate at which the system dynamics is happening, while the nonlinear contribution is not. Then, we can perform a rotating wave approximation on the time-variation in this equation which introduces a delta function in the summation:
\begin{align}
\sum_{\mu, \nu, \alpha, \beta} \Lambda_{\mu, \nu, \alpha, \beta}A_\alpha^*(t) A_\nu(t) A_\beta(t) e^{i(\omega_\alpha + \omega_\mu - \omega_\nu - \omega_\beta)t} \approx  e^{i\delta_\mu t}\sum_{ \nu, \alpha, \beta} \Lambda_{\mu, \nu, \alpha, \beta}\delta_{\mu + \alpha, \nu + \beta}a_\alpha^*(t) a_\nu(t) a_\beta(t)
\end{align}
where we have introduced $a_\mu(t) = A_\mu(t) e^{-i\delta_\mu t}$. Substituting for $\Lambda_{\mu, \nu, \alpha, \beta}$, we obtain:
\begin{align}
&\sum_{ \nu, \alpha, \beta}\Lambda_{\mu, \nu, \alpha, \beta}A_\alpha^*(t) A_\nu(t) A_\beta(t) e^{i(\omega_\alpha + \omega_\mu - \omega_\nu - \omega_\beta)t} \approx e^{i\delta_\mu t} \bigg[\sum_{\nu, \alpha, \beta} a_{\nu + \beta - \mu}^* a_{\nu}a_{\beta} + 2 a_\mu \sum_{\alpha} |a_\alpha|^2\bigg]
\end{align}
We can thus rewrite Eq.~\ref{eq:main_dynamics} as:
\begin{align}
\boxed{
\frac{da_\mu(t)}{dt} = -\bigg[i\delta_\mu + \frac{\kappa_\mu + \gamma_\mu N(t)}{2}\bigg] a_\mu(t) + ig_0 (1 + i\xi) \bigg[\sum_{\nu, \beta} a_{\nu + \beta - \mu}^* a_{\nu}a_{\beta} + 2 a_\mu \sum_{\alpha} |a_\alpha|^2\bigg] + \sqrt{\kappa_0 \eta}s_{\text{in}}\delta_{\mu, 0}}
\end{align}
where $g_0 = -\Lambda_0  (i\sigma_{\text{TPA}} + \varepsilon_0 \omega_0 \chi) / 2\varepsilon_0 V_0$. Similarly simplifying the nonlinear term in Eq.~\ref{eq:carrier_dynamics}:
\begin{align}
\sum_{\mu, \alpha, \nu, \beta}\Lambda_{\mu, \nu, \alpha, \beta} A_\mu^* A_\alpha^* A_\nu A_\beta e^{i(\omega_\mu + \omega_\alpha - \omega_\nu -\omega_\beta)t} \approx \Lambda_0 \bigg[\sum_{\mu, \nu, \beta} a_\mu^* a_{\nu + \beta - \mu}^* a_\nu a_\beta + 2\bigg(\sum_{\mu}|a_\mu|^2 \bigg)^2\bigg]
\end{align}
yielding
\begin{align}
\boxed{
\frac{dN(t)}{dt} = -\frac{N(t)}{\tau_d} + g_0 \xi_0 \bigg[\sum_{\mu, \nu, \beta} a_\mu^* a_{\nu + \beta - \mu}^* a_\nu a_\beta + 2\bigg(\sum_{\mu}|a_\mu|^2 \bigg)^2\bigg]}
\end{align}
Next, we put these equations in a dimensionless form for ease of analysis. We introduce the pump threshold power (in photons per second) $p_{\text{th}}$ ignoring TPA:
\begin{align}
p_{\text{th}} = \frac{\kappa^2}{8g_0\eta}
\end{align}
where $\kappa$ is the average decay rate of the resonators. Introducing normalized mode amplitudes $\alpha_\mu = a_\mu \sqrt{2g_0 / \kappa}  $ and normalized time $\tau = \kappa t / 2$, we obtain:
\begin{align}
\frac{d\alpha_\mu(\tau)}{d \tau} = -(\lambda_\mu + \theta_\mu N(\tau))\alpha_\mu(\tau) + (i - \xi)\bigg[\sum_{\nu,\beta}\alpha_{\nu + \beta - \mu}^*(\tau) \alpha_\nu(\tau) \alpha_\beta(\tau) + 2 \alpha_\mu(\tau) \bigg(\sum_\nu |\alpha_\nu(\tau)|^2\bigg) \bigg] + \delta_{\mu, 0} f_{\text{in}}(\tau)
\end{align}
  where $\lambda_\mu = (2i \delta_\mu + \kappa_\mu) / \kappa$ and $\theta_\mu = \gamma_\mu / \kappa$. Similarly:
  \begin{align}
  \frac{dN(\tau)}{d\tau} = -\lambda_d N(\tau) + \xi_N \bigg[\sum_{\mu, \nu, \beta} \alpha_\mu^* \alpha_{\nu + \beta - \mu}^* \alpha_\nu 
 \alpha_\beta + 2\bigg(\sum_{\mu}|\alpha_\mu|^2 \bigg)^2\bigg]
  \end{align}
  where $\lambda_d = 2 / \kappa \tau_d$ and $\xi_N = \kappa \xi / 2g_0$.

\end{document}